\documentclass[notitlepage,aps,prx,twoside,superscriptaddress,showkeys,nofootinbib,11pt]{revtex4-1}
\usepackage{graphicx}
\usepackage{amsmath,amsfonts,amssymb,amsthm,amsbsy,mathtools}
\usepackage{bbold}
\usepackage{epic}
\usepackage{eepic}
\usepackage{mathrsfs}
\usepackage[all]{xy}
\usepackage{lmodern}
\usepackage[T1]{fontenc}
\usepackage[utf8]{inputenc}
\usepackage{color}
\usepackage{xcolor}
\usepackage{centernot}
\usepackage{mathtools}
\usepackage{ stmaryrd }
\usepackage{palatino}

\usepackage[breaklinks=true,colorlinks=true,linkcolor=blue,urlcolor=blue,citecolor=blue]{hyperref}

\newmuskip\pFqmuskip   

\newcommand*\pFq[6][8]{%
  \begingroup 
  \pFqmuskip=#1mu\relax
  \mathcode`\,=\string"8000
  \begingroup\lccode`\~=`\,
  \lowercase{\endgroup\let~}\pFqcomma
  {}_{#2}F_{#3}{\left[\genfrac..{0pt}{}{#4}{#5};#6\right]}
  \endgroup
}
\newcommand{\pFqcomma}{\mskip\pFqmuskip}

\begin{document}
	
\title{Comment on ``Revisiting dynamics of quantum causal structures -- when can causal order evolve?''}

\author{Esteban Castro-Ruiz}
\affiliation{Institute for Quantum Optics and Quantum Information (IQOQI),
Austrian Academy of Sciences, Boltzmanngasse 3, A-1090 Vienna, Austria}
\affiliation{QuIC, Ecole polytechnique de Bruxelles, C.P. 165,
Universit\'e libre de Bruxelles, 1050 Brussels, Belgium}
\author{Flaminia Giacomini}
\affiliation{Perimeter Institute for Theoretical Physics, 31 Caroline St. N, Waterloo, Ontario, N2L 2Y5, Canada}
\author{\v{C}aslav Brukner}
\affiliation{Institute for Quantum Optics and Quantum Information (IQOQI),
Austrian Academy of Sciences, Boltzmanngasse 3, A-1090 Vienna, Austria}
\affiliation{Vienna Center for Quantum Science and Technology (VCQ), Faculty of Physics,
University of Vienna, Boltzmanngasse 5, A-1090 Vienna, Austria}

\begin{abstract}
\noindent  In the last few years, there has been increasing interest in quantum processes with indefinite causal order. Process matrices are a convenient framework  to study such processes. Ref.~\cite{Dynamics} defines higher order transformations from process matrices to process matrices and shows that no continuous and reversible transformation can change the causal order of a process matrix. Ref.~\cite{Selby} argues, based on a set of examples, that there are situations where a process can change its causal order over time. Ref.~\cite{Selby} claims that the formalism of higher order transformations is not general enough to capture its examples. Here we show that this claim is incorrect. Moreover,  a crucial example of Ref.~\cite{Selby} has already been explicitly considered in Ref.~\cite{Dynamics} and shown to be compatible with its results. 
\end{abstract}
	
\maketitle


\noindent Ref.~\cite{Selby} considers 4 examples. The first 2 involve the change of quantum channels over time and are conceptually equivalent. Both introduce a fibre along which a quantum system $S$ propagates. If the fibre is perfect, the quantum channel implemented is the identity on $\mathcal{L}(\mathcal{H}_S)$, the Hilbert space of linear operators corresponding to $S$. However, the fibre can stop working due to some microscopic effects in its components. If these effects are significant, the channel implemented can be maximally depolarising. Ref.~\cite{Selby} considers a scenario where the fibre starts off being perfect and then, after time  $t$, it changes  into a useless fibre. The whole process is reversible in time due to the unitarity of quantum mechanics. 

The authors claim that this scenario cannot be captured by a higher order transformation $A(t)$ acting on $\mathcal{L}(\mathcal{L}(\mathcal{H}_S))$ such that $C(t) = A(t)[C(0)]$, where $C(0)$ is the identity on  $\mathcal{L}(\mathcal{H}_S) $ and $C(t)$ is the maximally depolarising channel on  $\mathcal{L}(\mathcal{H}_S) $. This claim is correct. However, the claim that this scenario cannot be captured by the framework of higher order transformations is incorrect. In fact, this scenario can clearly be modelled by a higher order transformation $\tilde{A}(t): \mathcal{L}(\mathcal{L}(\mathcal{H}_S\otimes\mathcal{H}_F)) \longrightarrow \mathcal{L}(\mathcal{L}(\mathcal{H}_S))$ such that $C(t) = \tilde{A}(t)[\tilde{C}]$, where $\tilde{C}$ is a channel on \textit{both} the system and the microscopic degrees of freedom of the fibre, whose Hilbert space we denote by $\mathcal{H}_F$. $\tilde{C}$ is such that $\tilde{A}(0)[\tilde{C}] = C(0)$ (the identity on $\mathcal{L}(\mathcal{H}_S)$) and $\tilde{A}(t)[\tilde{C}] = C(t)$ (the maximally depolarising channel on $\mathcal{L}(\mathcal{H}_S)$). Although the internal microscopic dynamics of the fibre is continuous and reversible as a function of $t$, the map $\tilde{A}(t)$ is not reversible, because the Hilbert space of its domain has a higher dimensionality than that of the Hilbert space of its range. These type of transformations were introduced in Section VD of~\cite{Dynamics} in a general way.  

The third example concerns a fibre that ``connects'' the local laboratory of $A$ to that of $B$. Ref. ~\cite{Selby} considers a scenario where the fibre’s microscopic constituents continuously evolve over time $t$ in such a way that the fibre changes from the configuration correspoding to the channel from A to B, to the one corresponding to the channel from B to A. At initial time $t=0$, the fibre configuration implements a channel $C_{A\rightarrow B}: \mathcal{L}(\mathcal{H}_A) \longrightarrow \mathcal{L}(\mathcal{H}_B)$. At a later time $t$, the fibre implements a channel  $C_{B\rightarrow A}: \mathcal{L}(\mathcal{H}_B) \longrightarrow \mathcal{L}(\mathcal{H}_A)$. It is true that this scenario does not correspond to a transformation $A(t)$ such that $C_{B\rightarrow A} = A(t)[C_{A\rightarrow B}]$. However, it is clear that this scenario, as well as the previous examples, can be modelled by a higher order transformation $\tilde{A}(t): \mathcal{L}(\mathcal{L}(\mathcal{H}_A\otimes \mathcal{H}_F )\longrightarrow \mathcal{L}(\mathcal{L}(\mathcal{H}_B))$ such that $\tilde{A}(0)[\tilde{C}] = C_{A\rightarrow B}$ and $\tilde{A}(t)[\tilde{C}] = C_{B\rightarrow A}$  for some $\tilde{C}$ on $\mathcal{L}(\mathcal{H}_A\otimes\mathcal{H}_F)$. Again, $\tilde{A}$ is not reversible as a higher order transformation, even if the constituents of the fibre evolve unitarily in time. 

The last example concerns the so-called quantum switch~\cite{Chiribella}: a quantum process in which the order of signaling between $A$ and $B$ is coherently controlled by an extra quantum system. The authors consider a scenario where, following the same logic of the previous examples, one can map the quantum switch into a process with definite causal order (a channel from $A$ to $B$, say). This example is explicitly constructed in Section VD of Ref.~\cite{Dynamics}, where it was argued that continuous and reversible local operations in the past \textit{can} influence the causal order of local operations in the future. Surprisingly, Ref.~\cite{Selby} does not comment on this example. More generally, Ref.~\cite{Selby} makes no reference to Section VD of~\cite{Dynamics}, which includes all their examples and their approach to channel dynamics as particular cases. Instead, Ref.~\cite{Selby} claims its last example ``is another instance where the lack of generality in the scope of the framework of higher-order processes is evidenced". This claim is evidently false, as the example Ref.~\cite{Selby} sketches was explicitly constructed in Ref.~\cite{Dynamics} \textit{using} the framework of higher-order processes.

The conclusion of Ref.~\cite{Dynamics} that the causal order cannot be changed by a continuous and reversible transformation was reached by studying transformations that map the \emph{an arbitrary} process matrix to a valid process matrix. This is a crucial requirement, as a transformation which might be allowed when applied to a specific process matrix might map a different process matrix out of the set of allowed process matrices, and hence would not be allowed as a process matrix transformation. If this happens, we either need to restrict the domain of the set of the transformations (and hence we would not be considering all the causal structures that are mathematically allowed) or we are looking at a restricted scenario, which does not allow us to derive any general statement about the dynamics of the causal structure. This is a methodological disadvantage of the approach taken in Ref.~\cite{Selby}. For this reason, we have opted to use the more advantageous approach of  Section VD in Ref.~\cite{Dynamics}: on the one hand, it can model all the type of situations considered in Ref.~\cite{Selby}. On the other hand, it is perfectly compatible with the framework of higher-order maps.  

When assessing the origin of the apparent contradiction between~\cite{Dynamics} and~\cite{Selby}, Ref.~\cite{Selby} claims that ``we can refine the statement of the result of Ref. [21] [Ref.~\cite{Dynamics} in this manuscript] as follows: \textit{Predictable}, reversible and continuous dynamics of \textit{black-box} process-matrices cannot change causal order.'' According to Ref.~\cite{Selby}, ``Predictable'' means that one cannot influence the inner workings of the physical system that implements the channel, like for example moving beamsplitters around in an optical table. ``Black box'' means that ``the only information we have access to is the CP map associated to a channel or process-matrix''. We now clarify these notions with respect to those used in Ref.~\cite{Dynamics}.  In Ref.~\cite{Dynamics} it is clearly stated that process matrices are objects with no ``open-ends". According to Ref.~\cite{Dynamics}, an object has no open ends if it forms a closed system, where ``the probabilities for the events described in the laboratories are completely determined by the choice of local operations performed by the parties and the way the process matrix connects the laboratories''. (This condition was introduced in \cite{Hardy} for the case of quantum circuits.) From these definitions it follows that ``processes with no open ends'' are ``predictable black boxes''. This reasoning shows that Ref.~\cite{Selby} does not provide any ``refinement'' of the results in \cite{Dynamics}. On the contrary, it shows that the assessment of ~\cite{Dynamics} presented in~\cite{Selby} is incomplete. 

Finally, note that demanding that processes have no open ends is a prerequisite so that we can unambiguously say a process has certain (definite or indefinite) causal order. With open ends, the direction of signaling between $A$ and $B$ might depend on how we ``close'' the open ends. This condition is particularly important in the case of indefinite causal structures, where the physical mechanism implementing a process might not be fully understood. In view of this observation,  one can state the theorem in Ref.~\cite{Dynamics} in the following, operational way: Imagine a set of local laboratories (parties) are ``connected'' in a specific way, denoted by $W$. Suppose $W$ undergoes a transformation $A$ such that the signalling correlations between the parties change. For example, it could happen that the parties are not able to violate a causal inequality (see \cite{OCB}) before $A$ but are able to do so after $A$. Then, at least one of the following statements is true: \textit{i)} $A$ is  discontinuous; \textit{ii)} $A$ is irreversible; \textit{iii)}  $W$  has open ends. The latter means  there are relevant extra physical variables that we need to take into account. 

In conclusion, the apparent contradiction between~\cite{Selby} and~\cite{Dynamics} is due to the fact that the assessment of~\cite{Dynamics} presented in~\cite{Selby} is incomplete.


\section*{Acknowledgements}
We thank J. Selby, A.B. Sainz and P. Horodecki for discussions.

\end{document}